# Electrostatic solitary waves in ion beam neutralization

C. Lan[1,2] and I. D. Kaganovich[1]

[1] *Princeton Plasma Physics Laboratory, Princeton, New Jersey 08543, USA*
[2] *Institute of Fluid Physics, China Academy of Engineering Physics, Mianyang, 621900, P. R. China.*

Abstract: The excitation and propagation of electrostatic solitary waves (ESWs) are observed in two-dimensional particle-in-cell simulations of ion beam neutralization by electron injection by a filament. Electrons from the filament are attracted by positive ions and bounce inside ion beam pulse. Bouncing back and forth electron streams then starts to mix, creating the two-stream instability. The instability saturates with the formation of ESWs. These ESWs reach several cms in longitudinal size and are stable for a long time ($\gg \tau_b$, the duration of the ion beam pulse). The excitation of large-amplitude ESWs reduces the degree of neutralization of the ion beam pulse. In addition, the dissipation of ESWs causes heating of neutralizing electrons and their escape from the ion beam, leading to further reduction of neutralization degree.

Ion beam neutralization has attracted much attention in the past few decades and finds applications in many fields involving astrophysics [1], accelerator applications [2], inertial fusion, in particular fast ignition [3] and heavy ion fusion [4], as well as ion beam based surface engineering [5]. In laboratory, these electrons can come from gas ionization caused by ion beam, hot emission of filament, secondary electron emission caused by ion bombardment on metal or pre-formed plasma in the channel of ion beam propagation. Because of the mass disparity between ions and electrons, this process can occur rapidly on the time scale of intense pulsed beams. As a typical application, heavy ion fusion research facility like NDCX requires near-complete (99%) space charge neutralization [6, 29]. Incomplete neutralization results in transverse emittance growth, even defocusing of the ion beam. In ion beam etching or nanopantography, space charge compensation is particularly important for ion beams with high current density [5]. One of central issues of ion beam neutralization is to determine what factors would affect the ultimate degree of charge and current neutralization. In the past two decades, it has been found that these factors include solenoidal magnetic field [7], large-amplitude plasma wave excitation [8, 9], and the way of electron supply [6], etc. In this Letter, we show that a new mechanism, i.e. the excitation of nonlinear electrostatic solitary waves (ESWs), can essentially affect the capture of electrons and the degree of ion beam neutralization.

ESWs were originally discovered during simulations of the nonlinear stage of the two-stream instability [10], and have been observed in space and laboratory plasmas for many years [11-19]. See Hutchinson's paper for a recent review of ESWs [20]. But surprisingly, as far as we know they have never been reported in ion beam neutralization experiments, and even never been mentioned or investigated in related literatures. In those space observations, ESWs are usually positive potential structures moving along the ambient magnetic field with typical speed on the order of 1000 km/s and duration on the order of 10 ms [21]. In phase space, in order to support a positive potential structure, trapped electrons form a localized hole. So ESWs are also often referred to by other related terms such as electron hole [22-25]. For ion beam neutralization, decreased electron density results in a local maximum in charge density and hence in electric potential. The effect of solitons on the degree of ion beam neutralization and the transverse movement of ions can not be ignored.

ESWs are essentially a type of Bernstein-Greene-Kruskal (BGK) mode [26] and do not experience any decay in one-dimensional plasma. However, the existence and stability criteria for multi-dimensional ESWs in unmagnetized plasmas are still under studied. In Ref. [20], it was summarized that for the existence of multi-dimensional ESWS, "there must be a strong enough magnetic field and distribution-function anisotropy". In unmagnetized plasmas, Ng and Bhattacharjee have shown that a distribution function that depends both on energy and angular momentum is also a possible condition [27]. Although ESWs can be formed in multi-dimension, they unavoidably experience transverse modulation and break-up [20, 28]. For the propagation and neutralization of ion beam pulse in a channel or a chamber, problems are at least two dimensional (2D). Whether ESWS can be excited and be even stable propagating in such multi-dimensional system is still unknown until now.

In the present Letter, we study by simulation the neutralization of a long ion beam pulse by electron injection. Past simulations of ESWs generally assume initial plasma is composed of two or more components of electrons with different velocity distributions [10, 21-22], which provides some default environments for the excitation of solitons. But here, these initial conditions for electrons are not necessary. The basic features of neutralization process, for instance, ion beam propagation and electron injection are preserved in our model. We investigate such a process using a 2D3V (2D in space but 3D in velocity) particle-in-cell code and show the behavior of electrons in ion beam pulse that has never been reported before.

We simulated long Gaussian-like $Ar^+$ ion beam pulses traveling in a metal pipe. The model is 2D in *x-y* cartisian coordinate system, where *x* is the traveling direction of ion beams and *y* is the transverse direction. The size of computational domain is 80 cm×3 cm. The energy and the maximal density of ion beam are $E_b$=38 keV, $n_b$=1.75×10$^{14}$ m$^{-3}$, respectively. The parameters of ion beam pulses were chosen to be close to those of Princeton Advanced Test Stand at PPPL [29]. Because beam velocity $V_b$ satisfies $V_b/c$<<1, where *c* is the speed of light in vacuum, for these parameters self-magnetic field can be neglected, so the computation is totally electrostatic. Ion beam pulses enter computational domain from left boundary and leave from right boundary. All the outer boundaries are absorbing boundaries for particles. Electrons were injected on the axis of the ion beam. Without loss of generality, the temperature of injected electrons is set to the typical temperature (~0.2 eV) of electrons emitted from a hot filament [30, 31]. Coulomb collisions were neglected as they only weakly affect the neutralization process. The cell size of uniform grid is Δ*l*=0.25 mm and the time step of simulations is Δ*t*=40 ps. The temperature of neutralizing electrons can be far higher than the initial temperature of electrons emitted from the filament.

Therefore, the Debye length $\lambda_d$ evaluated with the ion beam density and the temperature of neutralizing electrons satisfies $\Delta l < \lambda_d$.

When electrons are attracted by a positive ion beam pulse, electrons experience complex process in the potential well of ion beam pulse. Figure 1 shows how electrons interact with the potential well and are captured by the ion beam pulse. For the approaching space-charge potential well, downstream injected electrons are firstly accelerated and then reflected by the potential well. Some of these reflected electrons then once again are reflected by the other side of the potential well. Meanwhile, the potential of ion beam drops due to the filling of electrons into the potential well, leading to the escape of fast electrons and the capture of slow electrons. Thus, bouncing back and forth of trapped electrons naturally forms two streams in the potential well of the ion beam pulse, which will cause the occurrence of instability in phase space.

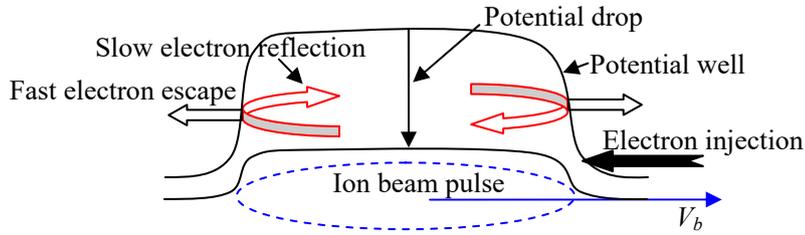

Fig. 1 Schematic of two-stream instability generation due to potential well of ion beam pulse. Dashed ellipse represents ion beam pulse. Electrons are injected downstream.

Our simulations confirmed the existence of above physical mechanism and the possibility of subsequent excitation of ESWs after two-stream instability evolving into nonlinear stage. The results are exhibited in Fig. 2.

Figure 2 shows temporal evolutions of electron and ion densities on the $x$ axis and corresponding kinetic variation of electrons in $x$-$v_x$ phase space during the neutralization. At $t$=635 ns, ion beam pulse has crossed the injection point where filament is located ($x$=20 cm). The zigzag distribution of electron density and appearance of peaks at the edges of the ion beam pulse indicate that some injected electrons are reflected by the head and tail of ion beam pulse and two-stream instability is developing, although most electrons still pass through the ion beam pulse and are lost on the left wall. With the ion beam pulse moving forward, at 707 ns we see that almost all of new-injected electrons are reflected by the potential well and electron stream generates a big circle in the phase space. However, self-triggered two-stream instability leads to the mixing of electrons in the phase space and the split of this big electron hole into several smaller ones. The initial evolution stage of the two steam instability lasts until about 800 ns. In this stage, it is readily seen that some relatively small electron holes start to appear from $t$=779. Meanwhile, as electrons are continued to be injected, the neutralization degree of the ion beam pulse keeps increased.

When the two-stream instability evolves into second stage, we see that small electron holes tend to coalese and generate a large density hole (see *t*=995~1283 ns). After about a few hundred nanoseconds of coalescence, this large electron density hole becomes a stably solitary structure, namely ESW. In this period, beam neutralization basically saturates. Except for location where the solitary wave is, other parts of ion beam pulse reach near-complete neutralization.

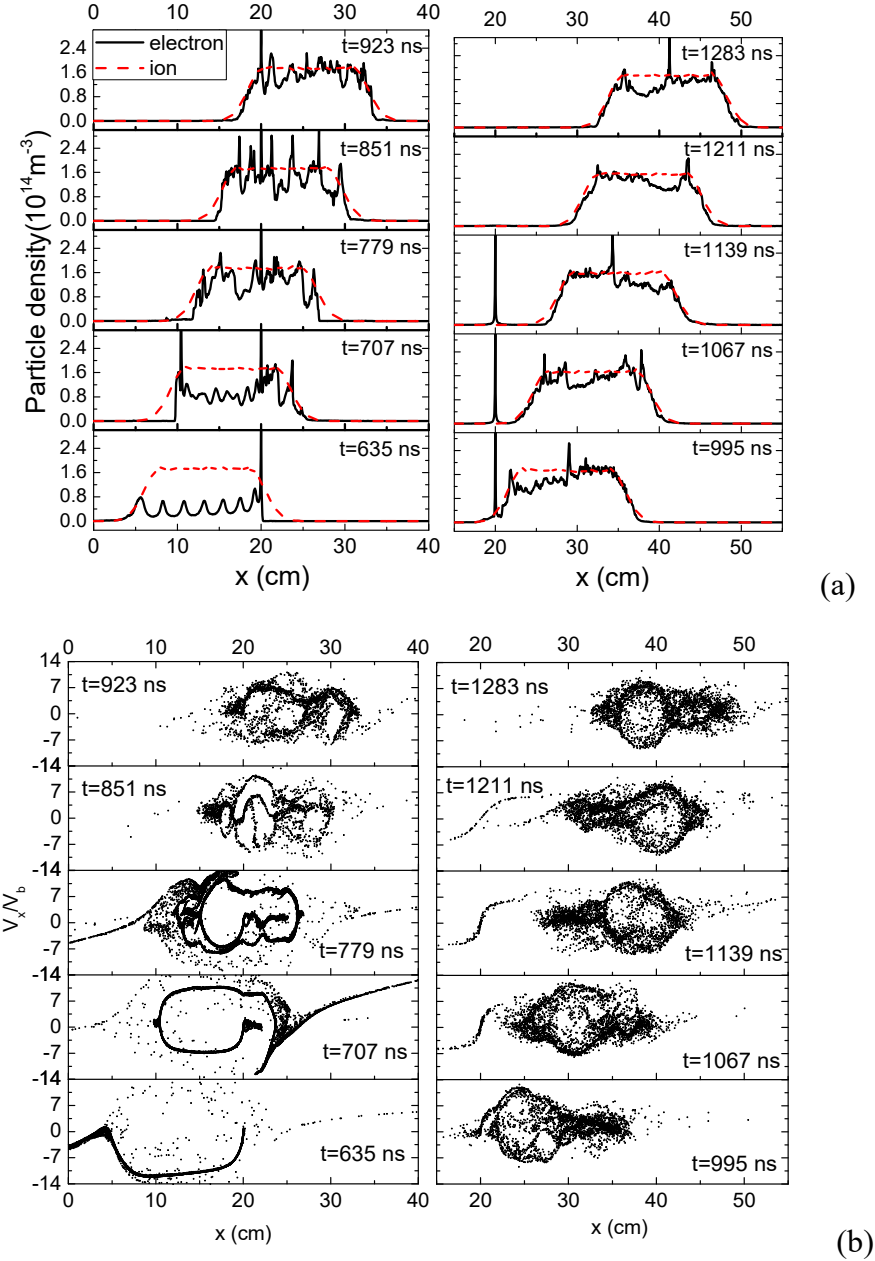

Fig .2 Temporal evolutions of particle densities on the axis (a) and electrons in $x$-$v_x$ phase space (b). The electron velocity is normalized to the ion beam velocity $V_b$=37.5 cm/μs. The simulated source of electron, which is located at $x$=20 cm, is a line source with 2 mm length in the $y$ direction. Electrons with $T_e$=0.2 eV are injected uniformly from 0 to 1.2 μs. The injected electron current $I_e$ is 2/3 of ion beam current $I_i$. The flattop of the ion beam is 250 ns (~11 cm). At the edges, the ion beam was taken to the form $n_b\exp(-t^2/t_b^2)$, where $n_b=1.75\times10^{14}$ m$^{-3}$, $t_b=60$ ns. The profile of the ion beam in the $y$ direction was taken to the from $n_b\exp(-y^2/w_b^2)$, where $w_b=2$ mm.

The coalescence process of small electron holes results in the formation of the ESW. To our surprise, the ESW moves rapidly inside the ion beam pulse, and when the ESW reaches the axial boundaries of the ion beam pulse, it can be reflected back with almost the same amplitude. As a result, the ESW moves back and forth periodically between the head and the tail of the ion beam pulse. In the phase space, the movement of the ESW is presented in a way of clockwise rotation (just like around a running track).

Fig. 3 shows variations of electron density and the potential on the axis when the ESW moves in different directions. The parameters of the ESW can be roughly estimated. In density space, the amplitude of the ESW reaches at least 2/5 of ion density and its longitudinal size reaches about 5 cm, close to 1/3 of ion beam length. The positive potential peak caused by localized density deficit of electron is about 20 V, that is 2/3 of residual ion beam potential. The period of the motion of the ESW is about 300 ns, close to the duration of the ion beam pulse. The speed of the ESW ($V_s$) relative to the ion beam pulse ($V_b=37.5$ cm/µs) is about 87 cm/µs (or 870 km/s), and the relative speeds in positive and negative directions are almost the same.

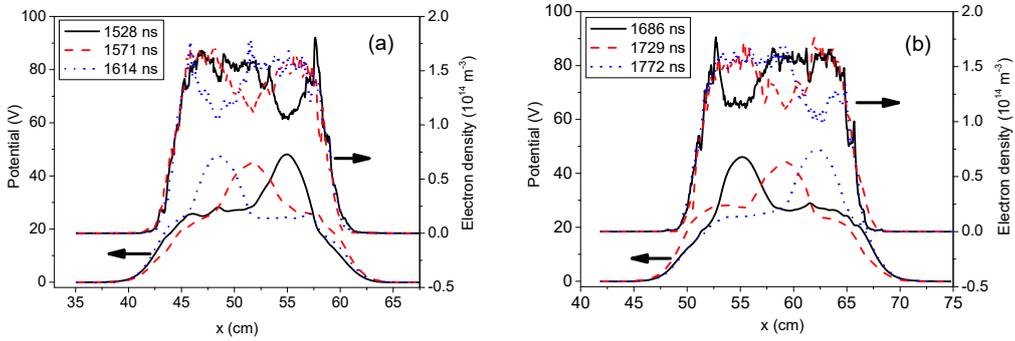

Fig. 3 Distributions of the electron densities and the potentials on the axis at different moments. Moving ion beam pulse is taken as a frame of reference. Two figures correspond to the propagation of the solitary wave to the left (a) and to the right (b), respectively.

In order to obtain the long lifetime of the ESW in a finite computational domain, periodic boundary conditions in the $x$ direction were applied. Then the ion beam pulse can reenter the computational domain after it exits from the right boundary. By recording time-dependent periodic voltage fluctuations of ion beam pulse at a certain point, the lifetime of the ESWs can be evaluated. In addition, to investigate effects of ESWs on the neutralization of the ion beam pulse, time-dependent variation of neutralization degree was also recorded, as shown in Fig. 4. The neutralization degree of the ion beam pulse is calculated by $\eta=Q_e/Q_i$, where $Q_e$ and $Q_i$ are the charges of electrons and ions in the whole computational domain, respectively.

It is readily seen from Fig. 4 that there are three periods of beam potential that have relatively intense ESW signals. After sixth period (5 μs), the ESW can not be visibly seen. So the propagation of the ESW lasts until 4~5 μs, and its lifetime is about 3.5 μs, which is much longer than the duration of the ion beam pulse.

The ESW would eventually decay to zero. The reason for the decay may be associated with the multi-dimension feature of the ESW. According to the existence and stability criteria for ESWs presented in Ref. [20], this 2D ESW generated in nonmagentized plasma is inherently unstable. But its long life still surprised us. Besides, the negative mass instability of trapped electrons in nonlinear plasma waves is another possible reason for the decay of the ESW [33, 34].

On the other hand, we see that because of the excitation of the ESWs, maximal value of $\eta$ only reaches 0.75 (solid line, $I_e/I_i=2/3$). Then it decays to a stable value (~0.64) with the attenuation of the ESW. The correlation between them indicates that the attenuation of the ESW could lead to gradual reduction of $\eta$. The reason for reduction is the heating of neutralizing electrons caused by the dissipation of the ESW. Once the energies of these thermalized electrons exceed the residual beam potential, they escape from the ion beam pulse. Therefore, it is concluded that *besides the excitation of the ESW, the attenuation of the ESW is another mechanism that causes the reduction of η.*

Figure 4 also shows the influence of injected current $I_e$ on $\eta$. When $I_e$ is increased to three times the original value, the neutralization of the ion beam pulse tends to saturate. However, from the attenuation of the neutralization degree, it is conjectured that the excitation of ESWs still occurs. So simply increasing the injected electron current can not completely eliminates the excitation of ESWs.

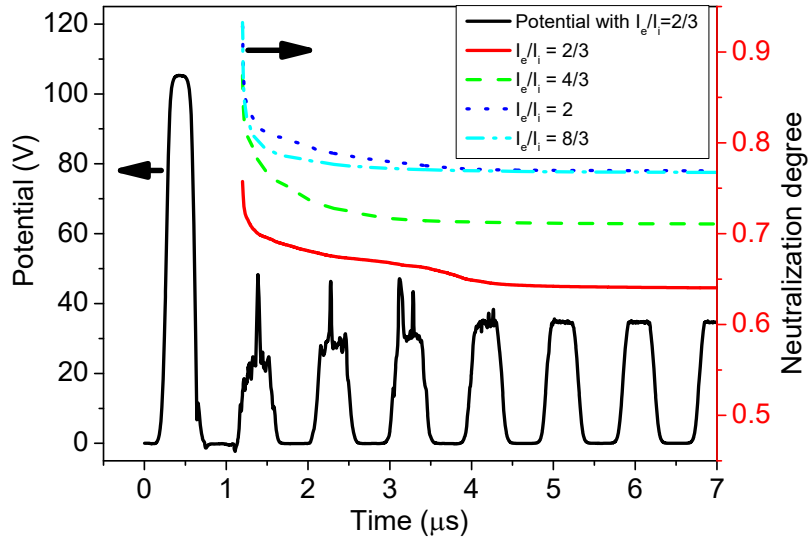

Fig. 4 Temporal evolutions of the potential recorded at *x*=5 cm on the axis and the neutralization degrees of the ion beam pulse for different injected electron currents. The size of computational domain is 40 cm×3 cm.

To achieve near complete neutralization of ion beam pulse, any excitation of ESWs and electron heating due to ESW-plasma interactions have to be minimized. One proposed method is to use a longitudinally extended electron source instead of a thin emitter. To prove the effectiveness of this method, we made electrons uniformly inject into an area ($\Delta x*\Delta y$=10 cm×2 mm), meanwhile injected current was increased to 4/3 of ion beam current. Figure 5 shows the simulation results. It is evident that the generation of electron holes is suppressed by injection of new electrons from downstream, and no any ESWs with amplitude comparable to that shown in Fig. 2 are created. The neutralization degree of the ion beam pulse finally reaches about 0.95, much higher than previous cases. However, more carefully observing the potential waveform, it can be found that ESWs still exist and even the number of solitary waves is more than 2. But their relative amplitudes are very small. Therefore, the affect of these solitary waves on the neutralization degree is not significant. In past experimental studies, it has been concluded that electrons produced through thermionic emission cannot neutralize ion beam well enough, and volumetric plasma can provide better charge neutralization for ion beam [5, 31]. It is reasonable to believe that the excitation of ESWs possibly is the main reason for that.

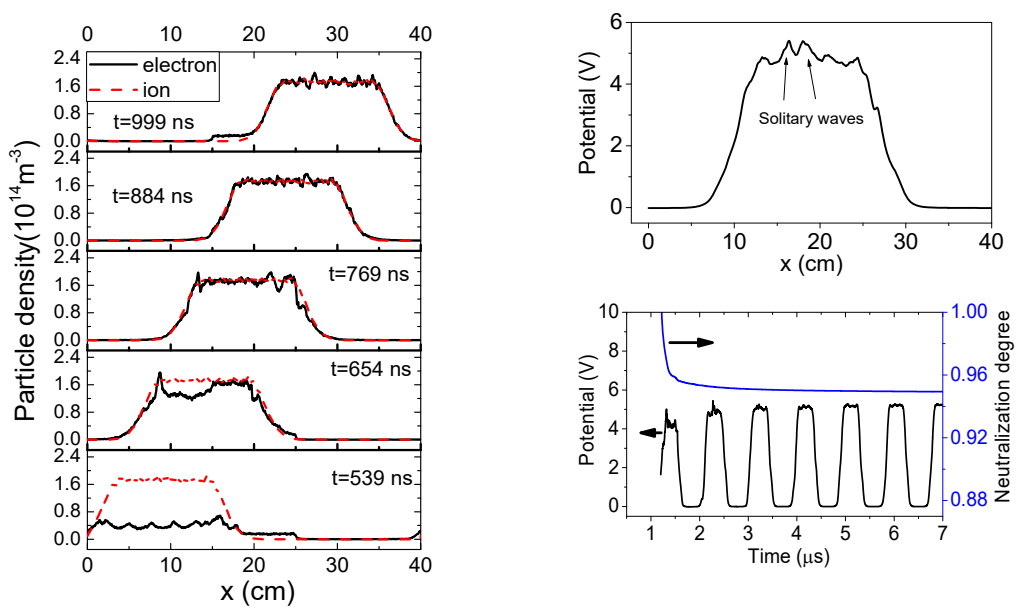

Fig. 5 Left: temporal evolutions of ion and electron densities on the axis. Upper right: beam potential on the axis at 2.04 μs. Right bottom: temporal evolutions of the neutralization degree of ion beam pulse and the potential recorded at $x$=5 cm on the axis. Electrons were injected uniformly from $x$=15 to 25 cm.

To summarize, we have presented 2D PIC simulation studies of the excitation and propagation of the ESWs during ion beam neutralization. We find that the capture process of electrons by ion beam pulse can cause the occurrence of the two-stream instability of electron, and more importantly this instability quickly evolves into stable moving nonlinear ESWs. The ESWs with longitudinal size reaching several cms can reflect back and forth between the two ends of the ion beam pulse for many times and last far longer than the duration of the ion beam pulse. The excitation of the ESWs reduces the neutralization degree of ion beam pulse. During the dissipation of ESWs, neutralizing electrons get energy from the ESWs and escape from the ion beam, leading to further reduction of neutralization degree. Simulations also show that simply increasing the injected electron current can not inhibit the excitation of the ESWs. But using a longitudinally extended electron source instead of a thin emitter can effectively minimize the excitation of the ESWs. These results provide a new insight in the physics of ion beam neutralization. In addition, our 2D model should help to study and understand the mechanisms of excitation and instability of BGK mode multi-dimensional solitons.

References:
[1] H. Alfven, Phys. Rev. **55**, 425 (1939); W. H. Bennett, Phys. Rev. **45**, 890 (1934); M.V. Medvedev and A. Loeb, Astrophys. J. **526**, 697 (1999); A. R. Bell, Mon. Not. R. Astron. Soc. **358**, 181 (2005).
[2] I. Blumenfeld *et al.*, Nature (London) **445**, 741 (2007); P. Chen *et al.*, Phys. Rev. Lett. **54**, 693 (1985); R. Govil *et al.*, Phys. Rev. Lett. **83**, 3202 (1999).
[3] A. J. Kemp *et al.*, Phys. Rev. Lett. **97**, 235001 (2006); R. J. Mason, Phys. Rev. Lett. **96**, 035001 (2006).
[4] P. K. Roy *et al.*, Phys. Rev. Lett. **95**, 234801 (2005); B. G. Logan *et al.*, Nucl. Instrum. Methods Phys. Res., Sect. A **577**, 1 (2007); I. D. Kaganovich *et al.*, *ibid.* **577**, 93 (2007); A. B. Sefkow *et al.*, *ibid.* **577**, 289 (2007); D. R. Welch, *ibid.* **577**, 231 (2007).
[5] J. P. Chang, J. C. Arnold, G. C. H. Zau, H.-S. Shin and H. H. Sawin, Journal of Vacuum Science & Technology A: Vacuum, Surfaces, and Films **15**, 1853-1863 (1997).
[6] I. D. Kaganovich, R. C. Davidson, M. A. Dorf, E. A. Startsev, A. B. Sefkow, E. P. Lee, and A. Friedman, Phys. Plasmas **17**, 056703 (2010).
[7] I. D. Kaganovich, E. A. Startsev, A. B. Sefkow, and R. C. Davidson, Phys. Rev. Lett. **99**, 235002 (2007).
[8] I. D. Kaganovich, E. Startsev, R.C. Davidson, Phys. Plasmas **11**, 3546 (2004).
[9] I. D. Kaganovich, A. B. Sefkow, E. A. Startsev, R. C. Davidson, and D. R. Welch, Nucl. Instrum. Methods Phys. Res. A **577**, 93 (2007).
[10] R. L. Morse and C. W. Nielson, Phys. Rev. Lett. **23**, 1087 (1969).
[11] K. Saeki *et al.*, Phys. Rev. Lett. **42**, 501 (1979).
[12] H. L. Pe´cseli *et al.*, Phys. Lett. **81**, 386A (1981); Phys. Scr. **29**, 241 (1984).
[13] H. L. Pe´cseli, Laser Part. Beams **5**, 211 (1987).
[14] G. Petraconi and H. S. Maciel, J. Phys. D **36**, 2798 (2003).
[15] M. Temerin, K. Cerny, W. Lotko, and F. S. Mozer, Phys. Rev. Lett. **48**, 1175 (1982).
[16] H. Matsumoto, H. Kojima, T. Miyatake, Y. Omura, M. Okada, I. Nagano, and M. Tsutsui, Geophys. Res. Lett. **21**, 2915 (1994).
[17] R. E. Ergun, C.W. Carlson, J. P. McFadden, F. S. Mozer, L. Muschietti, I. Roth, and R. J. Strangeway, Phys. Rev. Lett. **81**, 826 (1998).
[18] J. Franz, P. M. Kintner, and J. S. Pickett, Geophys. Res. Lett. **25**, 1277 (1998).

[19] C. A. Cattell, J. Dombeck, J. R. Wygnant, M. K. Hudson, F. S. Mozer, M.A. Temerin, W. K. Peterson, C. A. Kletzing, and C. T. Russell, Geophys. Res. Lett. **26**, 425 (1999).

[20] I. H. Hutchinson, Physics of Plasmas **24**, 055601 (2017).

[21] Q. M. Lu, D. Y. Wang, and S. Wang, J. Geophys. Res. **110**, A03223 (2005).

[22] H. L. Berk, C. E. Nielsen, and K.V. Roberts, Phys. Fluids **13**, 980 (1967).

[23] H. Schamel, Phys. Scr. **20**, 336 (1979).

[24] H. Schamel, Plasma Phys. **13**, 491 (1971); **14**, 905 (1972).

[25] H. Schamel, Phys. Plasmas **7**, 4831 (2000).

[26] I. B. Bernstein, J. M. Greene, and M. D. Kruskal, Phys. Rev. **108**, 546 (1957).

[27] C. S. Ng and A. Bhattacharjee, Phys. Rev. Lett. **95**, 245004 (2005).

[28] T. Miyake, Y. Omura, H. Matsumoto, and H. Kojima, J. Geophys. Res. **103**, 11841 (1998).

[29] A. D. Stepanov, E. P. Gilson, L. R. Grisham, I. D. Kaganovich, and R. C. Davidson, Physics of Plasmas **23**, 043113 (2016).

[30] D. V. Rose, D. R. Welch, S. A. MacLaren, Proceedings of the 2001 Particle Accelerator Conference, 3003 (2001).

[31] S. A. MacLaren, A. Faltens and P. A. Seidl, Phys. Plasmas **9**, 1712 (2002).

[32] I. D. Kaganovich, G. Shvets, E. A. Startsev, and R. C. Davidson, Phys. Plasmas **8**, 4180 (2001).

[33] I. Y. Dodin, P. F. Schmit, J. Rocks, and N. J. Fisch, Phys. Rev. Lett. **110**, 215006 (2013).

[34] K. Hara, T. Chapman, J. W. Banks, S. Brunner, I. Joseph, R. L. Berger, and I. D. Boyd, Phys. Plasmas **22**, 022104 (2015).